\documentclass{webofc}

\usepackage[varg]{txfonts}   
\usepackage{hyperref}
\usepackage{url}
\hypersetup{colorlinks=true,citecolor=blue,urlcolor=blue,linkcolor=blue}
\begin{document}
\title{Towards a fluid-dynamic description of an entire heavy-ion collision: from the colliding nuclei to	the quark-gluon plasma phase}
%
%

\author{\firstname{Andreas} \lastname{Kirchner}\inst{1}\fnsep\thanks{\email{andreas.kirchner@duke.edu}} \and
        \firstname{Federica} \lastname{Capellino}\inst{2} \and
        \firstname{Eduardo} \lastname{Grossi}\inst{3} \and
         \firstname{Stefan} \lastname{Floerchinger}\inst{3}
}

\institute{Department of Physics, Duke University, Durham, NC 27708, USA 
\and
           GSI Helmholtzzentrum für Schwerionenforschung, 64291 Darmstadt, Germany
\and
           Università di Firenze and INFN Sezione di Firenze, 50019 Sesto Fiorentino, Italy
           \and
           Friedrich-Schiller-Universität Jena, 07743 Jena, Germany
          }

\abstract{The fluid-dynamical modeling of a nuclear collision at high energy usually starts shortly after the collision. A major source of uncertainty comes from the detailed modeling of the initial state. While the collision itself likely involves far-from-equilibrium dynamics, it is not excluded that a fluid theory of second order can reasonably well describe its soft features. Here we explore this possibility and discuss how the state before the collision can be described in that setup, examine the required fluid-dynamical equations of motion and study the resulting entropy production. While we do here only first steps, we outline a larger program, whicht could lead to a dynamical description of heavy-ion collisions where the only uncertainty lies in the thermodynamic and transport properties of quantum chromodynamics.}
\maketitle
\section{Introduction}
\label{intro}
The fluid-dynamical modeling of heavy-ion collisions has been demonstrated to be a successful tool in enhancing our understanding of the underlying QCD dynamics of the collision and its subsequent dynamics \cite{Paquet:2023rfd,Heinz:2024jwu}. This great success raises the question of the range of applicability of fluid dynamics \cite{Schlichting:2024uok}. In this work we will explore the possibility of describing the soft part of the full collision purely by fluid dynamics, without any additional models for the initial collision. This description has the advantage of only requiring the equation of state and transport properties as inputs, which (in theory) all can be calculated from first principles.

We will begin this study by setting up the nuclei approaching each other in the beam pipe in terms of a fully fluid dynamic description, where the fluid fields of the composite system are obtained via Landau matching. Next we carry out the Landau matching procedure at different times to obtain the limit of a interactionless collision, allowing us to gauge the dynamics during the collision. As a final step we construct an equation of state spanning a large part of the QCD phase diagram and solve the fluid-dynamic equations for a drop of nuclear matter undergoing contraction and expansion. A more detailed explanation of the results presented here can be found in \cite{Kirchner:2024woh}.

\section{Nuclei as static fluids} \label{sec_staticFluids}

We begin by developing a fully fluid-dynamical description of the nuclei. This description is based on the conservation of energy, momentum and baryon number density current
\begin{align}
	\nabla_\mu T^{\mu\nu} =0, \quad \quad 	\nabla_\mu N^\mu =0. \label{eq:NumCons}
\end{align}
Hereby, the number density $n$ is modeled by a Woods-Saxon density profile. Since the nuclei traveling through the beam pipe before the collision are at the first order liquid-gas phase transition of the QCD phase diagram, their energy density can be easily related to their number density $\epsilon = \mu_\text{crit} n$ with $\mu_\text{crit} \approx 930 \; \text{MeV}$ being the critical chemical potential of the phase transition at $T=0$. The fluid velocity of a nucleus at rest is trivially given by $u^\mu=(1,0,0,0)$. With this the energy-momentum tensor and baryon number density current are obtained as
\begin{align}
	T^{\mu\nu} = \epsilon u^\mu u^\nu + p \Delta^{\mu\nu}, \quad N^\mu = n u^\mu. 
\end{align}
Since the nuclei are at the first order phase transition, the thermodynamic pressure is zero $p=0$, ensuring stability and preventing the nuclei from flowing apart. The collision system is then constructed by adding up two with opposite momentum boosted nuclei
\begin{align}
	T^{\mu\nu} = T^{\mu\nu}_\rightarrow +T_\leftarrow^{\mu\nu}, \quad
	N^\mu  = N^\mu_\rightarrow + N^\mu_\leftarrow. \label{eq:AddedN}
\end{align}
A description in terms of a single fluid is obtained by decomposing the energy-momentum tensor of the full collision system via Landau matching where the energy density is the eigenvalue of the time-like eigenvector of the energy-momentum tensor. This time-like eigenvalue is then the fluid velocity $T^\mu_\nu u^\nu = -\epsilon u^\mu$.

\section{Interactionless collision}
The interactionless limit is obtained by decomposing the composite system at different times $t$ according to the standard forms
\begin{align}
	T^{\mu\nu} = \epsilon u^\mu u^\nu + (p+\pi_\text{bulk})\Delta^{\mu\nu} + \pi^{\mu\nu}, \quad	N^\mu = n u^\mu + \nu^\mu,
\end{align}
with the nuclei being centered at $z_{\leftarrow/ \rightarrow} = \pm z_0 \mp vt$. Since there are no interactions in this matching procedure, the two nuclei pass through each other, displaying full transparency.
\begin{figure}[h]
	\centering
	\includegraphics[width=5.5cm,clip]{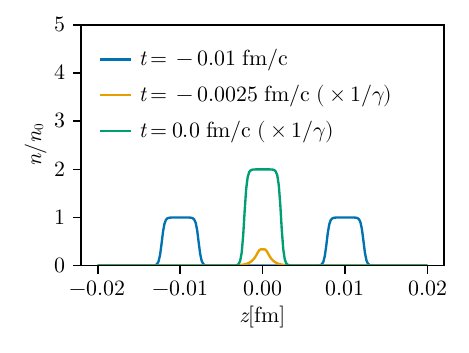}
	\includegraphics[width=5.5cm,clip]{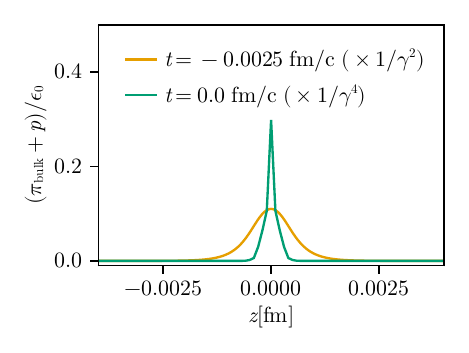}
	\caption{Number density (left) and pressure (right) for different times obtained from the Landau matching procedure in the interactionless limit. The large densities and viscous corrections require a careful treatment of the fluid-dynamical equations and their numerical implementation.}
	\label{fig_Density}       
\end{figure}
The resulting number density after the matching procedure can be seen in Fig.~\ref{fig_Density} (left panel), where the nuclei initially are well separated. During the overlap the density greatly increases, until it reaches its maximal value at full overlap. Due to symmetry, the dynamics after the overlap is the inverse of the dynamics before. Similarly, we also find a large increase in the viscous fields when the two nuclei begin to overlap (see Fig.~\ref{fig_Density}, right panel). These large densities and viscous fields then require a careful treatment of the evolution equations to ensure causality and numerical stability.

\section{Fluid dynamics \& equation of state}

The fluid dynamic description is based on the conservation laws Eq.~\ref{eq:NumCons}, together with the constituent relations for the viscous fields which are derived as proposed by Israel, Stewart and Muller \cite{Israel:1979wp,Muller:1967zza}, by demanding a positive entropy production at all space-time points.
The full derivation with the resulting equations can be found in \cite{Kirchner:2024woh}. Here we only want to concern us with two important points: Through the usage of second order viscous fluid dynamics additional transport coefficients have been introduced, with the main ones being the viscosities and conductivities ($\eta$, $\zeta$, $\kappa$) and their respective relaxation times ($\tau_\text{shear}$, $\tau_\text{bulk}$, $\tau_\text{heat}$).

These are relaxation type equations which allow the viscous fields to be away from their equilibrium value through the presence of relaxation times, ensuring that the description is valid outside of equilibrium. The second important point is that the relaxation times also allow for different levels of transparency in collision, as the limits of zero and infinite relaxation times result in full baryon stopping and full baryon transparency, respectively.

The equation of state is a combination of lattice QCD, the hadron resonance gas and a nucleon-meson model, in their respective areas of applicability (see \cite{Kirchner:2024woh} for details). With the equation of state we now have a complete set of equations together with a set of initial conditions that could be solved. However, as  discussed before, obtaining the solution of these equations is rather challenging due to causality and stability considerations. Therefore we will first examine a simplified system, which still captures a central aspect of a heavy-ion collision.

\section{Model system \& results}

\begin{figure}[h]
	\centering
	\includegraphics[width=5.5cm,clip]{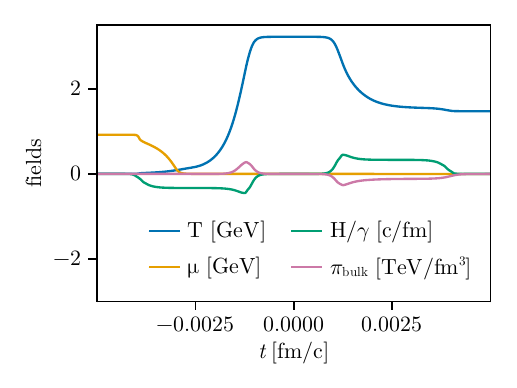}
	\includegraphics[width=5.5cm,clip]{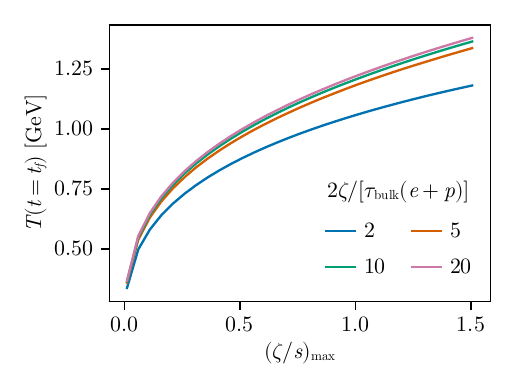}
	\caption{Fluid field evolution during the contraction and expansion (left) and final temperature as function of the viscosity (right). The system is initialized at zero temperature at the phase transition and ends its evolution with a finite temperature due to the produced entropy and heat.}
	\label{fig_OneEvent}       
\end{figure}

We will now solve the fluid-dynamic equations for a simplified system by focusing on the high compression and fast expansion of nuclear matter during the initial moments of a heavy-ion collision. This compression and expansion is modeled by filling a toy Hubble-like universe with nuclear matter and compressing and expanding it by rescaling all spatial dimensions with a time dependent scale factor 
	$\mathrm{d} s^2 = - \mathrm{d} t^2 + a(t)^2(\mathrm{d}x^2+\mathrm{d}y^2+\mathrm{d}z^2)$.
Since this universe is completely isotropic, homogeneous and invariant under rotations and translations, the only non-trivial fluid fields are the scalar quantities, depending only on time $\Phi(t)=(T,\mu,\pi_\text{bulk})$. We will model the Hubble rate $H=\dot{a}(t)/a(t)$ by employing the number density equation $\partial_t n + 3Hn=0$ and using the number density, obtained in the interactionless limit Fig.~\ref{fig_Density}. The resulting field solutions together with the Hubble rate can be seen in Fig.~\ref{fig_OneEvent}. Initially, the system sits at the phase transition with $T=0$ and $\mu=\mu_\text{crit}$. During the contraction of the system, the chemical potential decreases and the temperature increases. During the subsequent expansion the temperature decreases again, with its final value being larger than zero due to the entropy created through the viscosity of the system. The evolution of the bulk pressure initially shows a delay with respect to the Hubble rate, since it only begins to build up when entropy is being produced in equilibrium ($\mu \approx 0$).
The amount of produced entropy scales with the final temperature and therefore increases with increasing viscosity and decreasing relaxation time (Fig.~\ref{fig_OneEvent}, right panel). The final temperature is almost independent of the relaxation time, for small enough relaxation times, since the systems reaction to the forced contraction and expansion is almost instantaneous.
Since the final temperature is monotonously growing as function of the viscosity, we can choose a value of the viscosity together with the relaxation time to match any final temperature (e.g. the temperature found in the initial moments of the fireball).

\section{Conclusion}

In this work we have taken the first steps towards a fully fluid-dynamical simulation of the soft part of a heavy-ion collision by establishing the required equations of motion together with the equation of state. We solved these equations in a simplified system, modeling the compression and expansion of nuclear matter, allowing us to study the entropy production more closely. We are looking forward to combining our model with state of the art calculations of transport properties and their relaxation times, as well as studying the dynamics of the more complicated $1+1$D or $3+1$D systems.

%
%
%

\end{document}